\documentclass[aps,prl,floatfix,twocolumn]{revtex4}
 
\usepackage{graphicx}
\graphicspath{{figures/}}

\usepackage{amsmath}
\usepackage{SIunits}
\renewcommand{\vec}[1]{\boldsymbol{#1}}
\newcommand{\uvec}[1]{\hat{\boldsymbol{#1}}}
\newcommand{\abs}[1]{\left\vert #1 \right\vert}

\newcommand{\imaginary}[1]{\Im\left\{ #1 \right\}}

\newcommand{\vecr}{\!\left(\vec{r}\right)}
\newcommand{\vectorpotential}{\vec{A}(\vec{r},t)}
\newcommand{\amp}{u}
\newcommand{\amplitude}{\amp\vecr}
\newcommand{\intensity}{I\vecr}
\newcommand{\ph}{\varphi}
\newcommand{\phase}{\ph\vecr}
\newcommand{\pol}{\uvec{\epsilon}}
\newcommand{\polarization}{\pol\vecr}

\newcommand{\momentum}{\vec{g}\vecr}
\newcommand{\helicity}{\vec{\sigma}\vecr}
\newcommand{\spin}{\vec{s}\vecr}
\newcommand{\orbital}{\vec{\ell}\vecr}

\begin{document}

\title{Optical forces and torques in non-uniform beams of light}

\author{David B. Ruffner}
\author{David G. Grier}
\affiliation{Department of Physics and Center for Soft Matter Research,
New York University, New York, NY 10003}

\begin{abstract}
The spin angular momentum in
an elliptically polarized beam of light plays several noteworthy
roles in optical traps.
It contributes to the linear momentum density in a non-uniform
beam, and thus to the radiation pressure exerted on
illuminated objects. 
It can be converted into orbital angular momentum, and thus
can exert torques even on optically isotropic objects.
Its curl, moreover, contributes to both forces and torques
without spin-to-orbit conversion.
We demonstrate these effects experimentally by tracking 
colloidal spheres diffusing in 
elliptically polarized optical tweezers. 
Clusters of spheres circulate determinisitically
about the beam's axis.
A single sphere, by contrast, undergoes stochastic
Brownian vortex circulation that maps out the optical force field.
\end{abstract}


\maketitle

Optical forces arising from the polarization and
polarization gradients in vector beams of light constitute a 
new frontier for optical micromanipulation. 
Linearly polarized light has been used to 
orient birefringent objects in 
conventional optical tweezers 
\cite{friese_optical_1998,mohanty_optical_2005,preece_independent_2008}
and circular polarization 
has been used to make them rotate 
\cite{friese_optical_1998,oneil_three-dimensional_2000,%
oneil_intrinsic_2002,santamato_collective_1986,%
preece_independent_2008,bishop_optical_2004}.
More recently, optically isotropic objects also have been
observed to circulate in circularly polarized optical traps
\cite{zhao_spin--orbital_2007,zhao_direct_2009,wang_optical_2010},
through a process described as spin-to-orbit conversion
\cite{nieminen_angular_2008,wang_optical_2010,%
yan_comment_2011,wang_wang_2011,bomzon_angular_2006}.
Here, we present a general formulation of the linear and
angular momentum densities in vector beams of light
that clarifies how the amplitude, phase and polarization profiles
contribute to the forces and torques that such beams exert on 
illuminated objects.
This formulation reveals that the curl of the spin angular momentum
can exert torques on illuminated objects without contributing to the 
light's orbital angular momentum, and that this effect dominates
spin-to-orbit conversion in circularly polarized optical tweezers.
Predicted properties of polarization-dependent optical forces
are confirmed through observations of
a previously unreported mode of Brownian vortex circulation
for an isotropic sphere 
in elliptically polarized
optical tweezers.

The vector potential describing a beam of light of angular frequency
$\omega$ may be written as
\begin{equation}
  \label{eq:vectorpotential}
  \vectorpotential = \amplitude \, e^{i \, \phase - i \omega t} \, \polarization,
\end{equation}
where $\amplitude$ is the real-valued 
amplitude, $\phase$ is the
real-valued 
phase and $\polarization$ is the complex-valued
polarization vector at position $\vec{r}$.
This description is useful for practical applications
because $\amplitude$, $\phase$ and $\polarization$ may be
specified independently, for example using holographic techniques
\cite{grier03,roichman_optical_2008,preece_independent_2008,zhan_cylindrical_2009}.
Poynting's theorem then yields the time-averaged momentum density
\begin{equation}
  \label{eq:momentumdensitydefinition}
  \momentum
  = 
  \frac{\omega}{2 \mu c^2} \,
  \imaginary{
    \vec{A}^\ast(\vec{r},t) \times 
    \left[
      \nabla \times \vec{A}(\vec{r},t)
    \right]
  },
\end{equation}
where $\mu$ is the permeability of the medium
and $c$ is the speed of light in the medium.
The momentum density gives rise to the radiation
pressure that the light exerts on illuminated objects
and may be expressed in terms of the experimentally
accessible parameters 
as
\begin{equation}
  \label{eq:momentumdensity}
  \momentum
  = 
  \frac{\omega}{2 \mu c^2} \, 
  I\vecr \, \nabla \ph 
  - 
  \frac{i \omega}{2 \mu c^2} \, 
  I\vecr \, \epsilon^\ast_j \nabla \epsilon_j
  +
  \frac{1}{2}\nabla \times \vec{s},
\end{equation}
where $I\vecr = u^2\vecr$ is the intensity and where
\begin{equation}
  \label{eq:spin}
  \vec{s}\vecr
  =
  \frac{\omega}{2 \, \mu c^2} \, \intensity \, \helicity
  \end{equation}
is the spin angular momentum density in a beam of light
with local helicity
\begin{equation}
  \label{eq:helicity}
  \helicity = i \, \pol\vecr \times \pol^\ast\vecr.
\end{equation}
The projection of $\helicity$ onto the propagation direction
$\hat{k}\vecr$ is related 
to the Stokes parameters of the beam \cite{born_principles_1997} by
$\helicity \cdot \hat{k}\vecr = S_3\vecr / S_0\vecr$.
It achieves extremal values of $+1$ and $-1$ for right-
and left-circularly polarized light, respectively.

The momentum density described by Eq.~(\ref{eq:momentumdensity})
gives rise to the radiation pressure experienced by objects that
absorb or scatter light.
Identifying $\momentum$ with the radiation pressure on a particle
is most appropriate in the Rayleigh limit, when the particle's size is
no greater than the wavelength of light.
In this limit, the three terms in $\momentum$ may be interpreted as
distinct mechanisms by which a beam of light exerts forces
on illuminated objects.

The first two terms in Eq.~(\ref{eq:momentumdensity})
constitute the familiar phase-gradient contribution to
the radiation pressure \cite{roichman_optical_2008}.
In this context, the second term accounts for the independent
phase profiles that may be imposed on the real and imaginary
components of the polarization in an elliptically polarized beam.
Phase gradients have been used to create
three-dimensional optical force landscapes \cite{roichman_optical_2008}, 
such as knotted force fields \cite{shanblatt_extended_2011} 
and true tractor beams \cite{lee_optical_2010}.
They also account for the orbital angular momentum density
\begin{equation}
  \label{eq:orbital}
  \orbital = 
  \frac{\omega}{2 \mu c^2} \, \intensity \,
  \left[
    \vec{r} \times 
    \left( \nabla \varphi - i \epsilon_j^\ast \nabla \epsilon_j 
    \right) 
  \right],
\end{equation}
carried by
helical modes of light \cite{allen_orbital_1992,simpson_optical_1996}.
In this context, the polarization-dependent term in
Eq.~(\ref{eq:orbital}) vanishes identically in linearly polarized
light, but manifests spin-to-orbit conversion in elliptically
polarized beams.

\begin{figure*}[!t]
  \centering
  \includegraphics[width=0.98\textwidth]{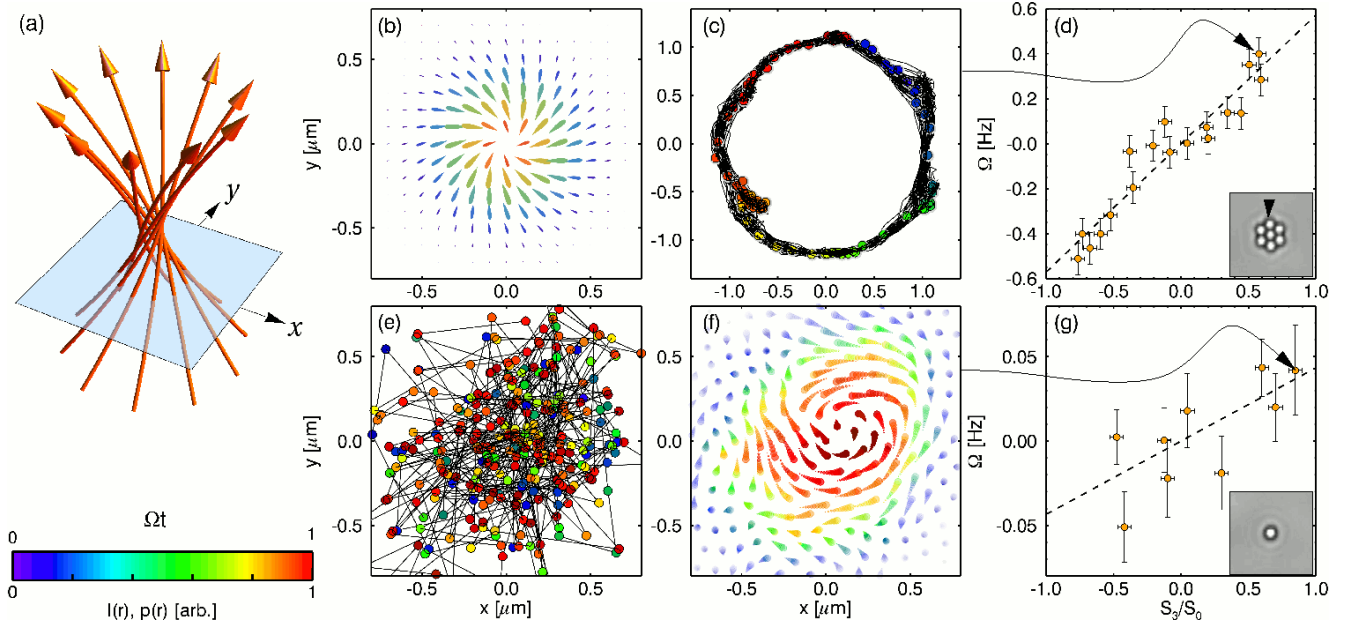}
  \caption{(Color online) (a) Streamlines of the momentum density 
    $\momentum$ in a 
    right-circularly polarized optical tweezer.
    (b) Components of $\momentum$ in the plane
    indicated in (a), shaded by the intensity $\intensity$.
    (c) Measured trajectory of one particle
    in a seven-sphere cluster trapped near the focus of the beam.
    Discrete points show the last three seconds of 
    motion, colored by time.
    (d) Circulation rate $\Omega$ as a function of the beam's
    Stokes parameters $S_3/S_0$.  Inset: snapshot of the cluster indicating
    the sphere whose trajectory is plotted.
    (e) Three seconds of a 3.5-minute trajectory of a single polystyrene sphere
    diffusing in a circularly polarized optical tweezer, shaded by time.
    (f) Time-averaged probability flux $\vec{j}(\vec{r})$ computed
    from the full measured trajectory.  Barbs are colored
    by the relative probability density $p(\vec{r})$ computed from
    the same trajectory.  Brownian vortex circulation is apparent in
    the vorticity of $\vec{j}(\vec{r})$.
    (g) Dependence of the Brownian vortex circulation rate on $S_3/S_0$.
    Inset: snapshot of the trapped sphere.
    The color bar indicates relative intensity
    $\intensity$ for (b), time for (c) and (e), and relative probability
    $p(\vec{r})$ for (f).
  }
  \label{fig:forces}
\end{figure*}

The third term in Eq.~(\ref{eq:momentumdensity}) 
describes how variations in spin angular momentum
contribute to the linear momentum density in non-uniform beams
of light.
This spin-curl term encompasses forces due to spatially-varying
elliptical polarization and also those due to intensity variations
in elliptically polarized beams.
Streamlines of $\nabla \times \vec{s}$ naturally loop
around extrema in the beam's intensity.
Spin-curl forces thus tend to make illuminated objects
circulate in the
plane transverse to the direction of propagation.
Observations of colloidal spheres circulating in
beams of light with spatially-varying elliptical polarization
\cite{cipparrone_polarization_2010,wang_optical_2010,wang_wang_2011}
consequently have been interpreted as evidence that the curl of the polarization
contributes to the light's orbital angular momentum.
Equation~(\ref{eq:orbital}), however, makes clear that the spin-curl
contribution to $\momentum$ does not contribute in any way to $\orbital$.
For the same reason, observations of optically-induced circulation in uniformly 
circularly-polarized optical traps 
\cite{zhao_direct_2009,zhao_spin--orbital_2007}
need not imply spin-to-orbit conversion.

To illustrate these point, we consider the forces exerted 
on an optically isotropic colloidal sphere by elliptically
polarized optical tweezers.
We model the trap as
an Gaussian beam of wavenumber $k$
brought to a focus with convergence angle $\alpha$
by a lens of focal length $f$ and 
numerical aperture $NA = n_m \sin\alpha$ in a medium of
refractive index $n_m$.
The beam's initial polarization is
\begin{equation}
  \label{eq:tweezerpolarization}
  \polarization = \frac{1}{\sqrt{2}} \, \left(
    \uvec{x} + e^{i\delta} \, \uvec{y} \right),
\end{equation}
with a corresponding incident helicity $\sigma_0 = \sin \delta$ along
$\uvec{z}$.
The focused beam's vector potential may be expressed
in cylindrical coordinates $\vec{r} = (\rho,\phi,z)$ with the
Richards-Wolf integral formulation 
\cite{wolf_electromagnetic_1959,richards_electromagnetic_1959,%
bomzon_angular_2006},
\begin{multline}
  \label{eq:wolffields}
  \vec{A}(\vec{r}) 
  = 
  -i \, 
  \left[ A_0\vecr + A_2\vecr \right] \,
  \left( 
    \cos \phi + e^{i \delta} \, \sin \phi
  \right) \, \hat{\rho} \\
  - i \,
  \left[ 
    A_0\vecr - A_2\vecr 
  \right] \,
  \left(
    e^{i\delta} \, \cos \phi - \sin \phi
  \right) \, \hat{\phi} \\
  - 2 \, A_1\vecr \, 
  \left(
    \cos \phi + e^{i \delta}  \, \sin \phi
  \right) \, \hat{z},
\end{multline}
as a Fourier-Bessel expansion
\begin{equation}
  \label{eq:An}
  A_n\vecr 
  = 
  \frac{k f u_0}{2 i\omega} \,
  \int_0^{\alpha} 
   a_n(\theta) \,
  J_n(k \rho \sin \theta) \,
  e^{i z k \cos \theta} 
  \, d\theta,
\end{equation}
with expansion coefficients \cite{richards_electromagnetic_1959}
\begin{align}
  \label{eq:expansioncoefficients0}
  a_0(\theta) & = 
  \left( 1 + \cos \theta \right) \, \sin\theta \sqrt{\cos \theta} \\
  \label{eq:expansioncoefficients1}
  a_1(\theta) & = \sin^2 \theta \, \sqrt{\cos \theta} \\
  \label{eq:expansioncoefficients2}
  a_2(\theta) & = 
  \left( 1 - \cos \theta \right) \, \sin\theta \, \sqrt{\cos \theta} .
\end{align}
Streamlines of $\momentum$ in
a right-circularly polarized optical tweezer ($\sigma_0 = +1$, $\text{NA} = 1.4$)
are shown spiraling 
around the optical axis in 
Fig.~\ref{fig:forces}(a).

A slice through the beam in the transverse 
plane indicated in Fig.~\ref{fig:forces}(a)
reveals the azimuthal component 
to the transverse momentum density
$g_\perp\vecr = \momentum \cdot \hat{\phi}$
that is plotted in Fig.~(\ref{fig:forces}b). 
The transverse momentum density may be resolved
into two contributions
\begin{equation}
  \label{eq:transversedensity}
  g_\perp\vecr = g_O\vecr + g_S\vecr
\end{equation}
arising from the spin-to-orbit and spin-curl
contributions to $\momentum$, respectively:
\begin{align}
  \label{eq:g_O}
  g_O\vecr 
  & =
  \frac{2 \omega}{\mu c^2} \, \frac{1}{\rho} \,
  \left[
    \abs{A_1\vecr}^2 + \abs{A_2\vecr}^2
  \right] \, \sigma_0 \quad \text{and}\\
  \label{eq:g_S}
  g_S\vecr & =
  \frac{\omega}{\mu c^2} \,
  \left[
    \partial_z \imaginary{A_1^\ast\vecr \, \left(A_0\vecr - A_2\vecr\right)} -
  \right. \nonumber \\
  & \quad\quad \left.
    \partial_r \left(
      \abs{A_0\vecr}^2 - \abs{A_2\vecr}^2
    \right) \right] \, \sigma_0.
\end{align}
Both 
are proportional to the
helicity of the incident beam, $\sigma_0$.
They do not, however, contribute equally to the transverse component of the
radiation pressure.
At the focus of the circularly-polarized optical tweezer,
for example, 79\% of the transverse momentum density is due to
the spin-curl term $g_S\vecr$ and only 21\% from spin-to-orbit
conversion.
More generally, 
both $A_1\vecr$ and $A_2\vecr$ vanish in the paraxial approximation;
there is no spin-to-orbit conversion in weakly focused beams.
The spin-curl contribution, by contrast, persists in the paraxial limit.

We probe the properties of spin-dependent optical forces
by measuring their influence on the motion of micrometer-scale
colloidal spheres. Our system consists of $1.0~\micro\metre$
diameter polystyrene (PS) spheres (Polysciences, Lot \# 586632)
dispersed in water and trapped in optical tweezers whose helicity
$\sigma_0$ is controlled with a quarter-wave plate.
The isotropic dielectric spheres absorb very little light
directly.
By scattering light, however, they experience radiation pressure
proportional to the local momentum density.
Our optical tweezer 
is powered by up to 4~\watt of laser light
at a vacuum wavelength of $\lambda = 532~\nano\metre$
(Coherent Verdi 5W).
The elliptically polarized beam is relayed with a dichroic mirror 
to the input pupil of an objective lens
(Nikon Plan Apo, $100\times$, NA 1.4), which focuses the light into a trap.
We account for the mirror's influence on the polarization by measuring
the beam's Stokes parameters in the input plane of the objective lens.
The sample is imaged using the same lens
in conventional bright-field
illumination, which passes through the dichroic mirror to a
video camera (NEC TI-324AII).
Digitally recorded video is analyzed 
with standard methods of digital video microscopy \cite{crocker96} to
measure the trajectory $\vec{r}_j = \vec{r}(j \tau)$
of a probe particle with 10~\nano\metre
resolution at $\tau = 33~\milli\second$ intervals.

The trajectory plotted in Fig.~\ref{fig:forces}(c) was 
obtained for one of seven spheres trapped against a glass surface
by a right-circularly-polarized optical tweezer ($\sigma_0 = +0.8$)
powered by 1.5~\watt.
The optically-assembled cluster, shown inset into Fig.~\ref{fig:forces}(d),
spans the region of the beam indicated in Fig.~\ref{fig:forces}(b),
and thus rotates about the beam axis at a rate of roughly 
$\Omega = 0.4~\hertz$.
The data in Fig.~\ref{fig:forces}(d) confirm the prediction of Eqs.~(\ref{eq:g_O})
and (\ref{eq:g_S}) that the rotation rate varies linearly with
the degree of circular polarization.

The colloidal cluster circulates deterministically in the elliptically polarized
optical tweezer because it continuously 
scatters light in regions where $g_\perp\vecr$ is substantial.
A single sphere diffusing in an elliptically polarized optical tweezer,
by contrast, explores the entire force landscape presented by the light.
This includes regions near the optical axis where $g_\perp\vecr$ is predicted
to vanish.
Figure~\ref{fig:forces}(e) shows the measured trajectory
of one such sphere in a right-circularly-polarized trap ($\sigma_0 = +0.8$)
powered by 0.05~\watt.
Optically-induced circulation is not immediately
obvious in the noisy trajectory, which is shaded to indicate the passage of time.
It becomes evident when the trajectory $\vec{r}_j$
is compiled into a time-averaged estimate \cite{silverman_density_1986}
for the steady-state probability current
\begin{equation}
  \label{eq:currentdenstiy}
  \vec{j}\vecr = \frac{1}{N-1} \, \sum_{j = 1}^{N-1}
  \frac{\vec{r}_{j+1} - \vec{r}_j}{\tau} \, 
  \delta_{\sigma_j}\!\left( 
    \vec{r} - \frac{\vec{r}_{j+1} + \vec{r}_j}{2} \right),
\end{equation}
which is plotted in Fig.~\ref{fig:forces}(f).
Here $N = 7,000$ is the number of discrete samples, and 
$\delta_\sigma\vecr$ is the kernel of an adaptive density estimator
\cite{silverman_density_1986}
whose width $\sigma$ varies with the sampling density. 
The symbols in Fig.~\ref{fig:forces}(f) are shaded by the
estimated probability density
\begin{equation}
  \label{eq:density}
  p\vecr = \frac{1}{N} \, \sum_{j = 1}^N
  \delta_{\sigma_j}\!\left(\vec{r} - \vec{r}_j\right)
\end{equation}
for finding the particle near $\vec{r}$.
Together, $\vec{j}\vecr$ and $p\vecr$ confirm the prediction of Eqs.~(\ref{eq:g_O})
and (\ref{eq:g_S}) that circulation vanishes on the optical axis where the
particle's probability density is greatest.

Taking care to measure $\vec{r}$ from the center of circulation, the mean
circulation rate may be estimated as
\begin{equation}
  \label{eq:circulationrate}
  \Omega 
  = 
  \int \rho\vecr \, 
  \left[ \vec{r} \times \vec{j}\vecr \right]
  \cdot \uvec{z} \, d^2r.
\end{equation}
Equation~(\ref{eq:circulationrate}) improves upon 
the graphical method for estimating $\Omega$
introduced in Ref.~\cite{roichman_influence_2008} by
making optimal use of discretely sampled data
\cite{silverman_density_1986}.
Because the single particle spends most of its time in a curl-free
region of the optical force field, its circulation rate is substantially
smaller than in the deterministic case.
Even so, the data in Fig.~\ref{fig:forces}(g) again are consistent
with the prediction that $\Omega$ scales linearly with $\sigma_0$.

The single particle's stochastic motion differs qualitatively
from the cluster's deterministic circulation.
Were it not for random thermal forces, the isolated sphere would
remain at mechanical equilibrium on the optical axis.
Thermal forces enable it to explore the optical force landscape, where
it is advected by the spin-dependent contribution to the radiation pressure.
This system, therefore constitutes an example of a Brownian vortex
\cite{sun_brownian_2009,sun_minimal_2010}, 
a stochastic machine that uses noise to transduce work out of a static
non-conservative force field.

Unlike previous experimental demonstrations of Brownian vortexes
\cite{roichman_influence_2008,sun_brownian_2009}
the conservative restoring force in this system
is transverse to the non-conservative contributions.
Consequently, the particle's radial excursions are described by
the Boltzmann distribution \cite{sun_minimal_2010}
$p\vecr = \exp\left(- \beta U\vecr\right)$
where $\beta^{-1} = k_B T$ is the thermal energy scale at
absolute temperature $T$ and 
$U\vecr$ is the particle's potential energy in the trap.
The solenoidal part of the radiation pressure \cite{sun_minimal_2010}
\begin{multline}
  \label{eq:nonconservative}
  \vec{f}\vecr = 
  \frac{\xi \omega}{8 \pi \mu c^2}  \, \nabla \times
  \int 
  \frac{\nabla^\prime \times I(\vec{r}^\prime) \left[
      \nabla^\prime \ph - i \pol_j^\ast \nabla^\prime \pol_j
    \right]}{\abs{\vec{r} - \vec{r}^\prime}} \, 
  d^3r^\prime \\
  + \frac{\xi}{2} \, \nabla \times \spin,
\end{multline}
then advects $p\vecr$ into the probability current
$\vec{j}\vecr = \mu_p p\vecr \, \vec{f}\vecr$,
where $\xi$ is the particle's scattering cross-section and $\mu_p$ is
its mobility.
The first term in Eq.~(\ref{eq:nonconservative}) may be neglected
in a beam such as a optical tweezer that
carries little or no orbital angular momentum.
The data in Fig.~\ref{fig:forces}(f) thus map out
the spin-curl force in the transverse plane.
Moreover, because the circulation direction is determined 
unambiguously by the curl of $\vec{f}\vecr$, this system
is a practical realization of a so-called
\emph{trivial Brownian vortex}, which has been proposed \cite{sun_minimal_2010}
but not previously demonstrated.

Formulating the optical momentum density in terms of experimentally
accessible parameters clarifies the nature and origin of the forces
that can be applied to microscopic objects using the radiation pressure
in beams of light.
This formulation confirms previous reports of forces arising
from phase gradients \cite{roichman_optical_2008} and demonstrates
that phase-gradient forces act independently of the state of polarization.
The spin-curl mechanism unifies 
forces arising from the curl of the polarization and forces due to
intensity gradients in elliptically polarized beams.
Because they induce circulatory motion,
spin-curl forces are easily misinterpreted as evidence for spin-to-orbit
conversion.
The spin-curl density, however, does not contribute to the orbital angular
momentum of the light.
Spin-to-orbit conversion, by contrast, removes spin angular momentum
from a beam of light and transmutes it into orbital angular momentum 
\cite{nieminen_angular_2008}.
The present formulation clarifies this mechanism, 
and reveals that spin-to-orbit conversion has played a secondary role
in previous reports of optically-induced circulation.
Using Eq.~(\ref{eq:momentumdensity}) as a guide, all three mechanisms
now may be optimally leveraged
to improve optical micromanipulation and the performance of light-driven machines.

We acknowledge helpful discussions with Giovanni 
Milione.
This work was supported by the
National Science Foundation principally through
Grant Number DMR-0855741, and in part by Grant Number DMR-0922680.


\end{document}